\documentclass[%
 reprint,
 amsmath,
 amssymb,
 aps,
pra,
]{revtex4-2}

\usepackage{graphicx}
\usepackage{dcolumn}
\usepackage{bm}
\usepackage{xcolor}
\usepackage{soul}

\usepackage{etoolbox}
\usepackage{physics}
\usepackage{qcircuit}
\usepackage{graphicx}
\usepackage{hyperref}
\usepackage{tikz}
\usepackage{amsmath}
\usepackage{algorithm}
\usepackage{algpseudocode}

\newcommand{\dnqv}{D$n$Q$v$ }

\begin{document}

\preprint{}

\title{Adaptive Lattice Gas Algorithm: \\Classical and Quantum implementations}

\author{Niccolò Fonio}
\altaffiliation[Also at ]{Laboratoir d'informatique et systems,Marseille}
\email{niccol.fonio11@gmail.com}
\author{Pierre Sagaut}%
 \altaffiliation[Also at ]{Quanscient Oy,Finland}
  \affiliation{%
 Aix Marseille Univ, Centrale Med, CNRS, M2P2 Laboratory UMR 7340, 13013 Marseille, France
}
%


\author{Ljubomir Budinski}
\altaffiliation[Also at ]{Faculty of Technical Sciences, University of Novi Sad, Serbia}
\author{Valtteri Lahtinen}
\affiliation{
 Quanscient Oy,Finland
}%


\date{\today}

\begin{abstract}
Lattice gas algorithms (LGA) are a class of algorithms including, in chronological order, binary lattice gas cellular automata (LGCA), integer lattice gas algorithms (ILGA) and lattice Boltzmann method (LBM).
They are largely used for simulating non-linear systems. Starting from 1-dimensional ILGA, we design an algorithm where we carry out a fraction of the possible collisions. These fractions are then adapted to reproduce LBM equilibrium distributions, resulting in an adaptive lattice gas algorithm (ALGA) that achieves the same simulation results of LBM. 
Considering this, we develop a quantum algorithm that involves a linear collision operator and capable of simulating the same phenomena, while still using a measurement and reinitialization procedure. Multi-time-step implementation is possible in some specific cases briefly discussed.
\end{abstract}

\maketitle


\section{Introduction} 

Lattice gas algorithms (LGA) include many numerical schemes mostly applied to simulating non-linear phenomena. They represent the evolution of a gas in discrete time-space using two steps: collision and streaming. The components of the system interact according to some conservation laws at each lattice point during the collision, changing their dynamical state. After the interaction, they are streamed according to their velocities. The components of a LGA can be microscopic particles, as in a lattice gas cellular automata (LGCA) \cite{Rivet_Boon_2001,wolf2004lattice,frisch2019lattice}, or mesoscopic probability density functions (PDF) as in the case of lattice Boltzman method (LBM) \cite{kruger2016lattice, mohamad2011lattice} \cite{succi2001lattice,succi2018lattice}. LGA being among the most used computational fluid dynamics (CFD) methods, an advantage in their quantum computing (QC) implementation is being investigated closely in the developing field of quantum computational fluid dynamics (QCFD) \cite{bharadwaj2020quantum}.

The intersection between LBM and quantum mechanics has been studied since the early '90s \cite{succi1993lattice}. Specifically, quantum LGCA \cite{meyer1996quantum,meyer1997quantum,boghosian1998quantum,yepez1998lattice, yepez2001quantum, yepez2001quantumcfd, fonio2023quantum, zamora2025efficient} and quantum LBM \cite{budinski2021quantum, ljubomir2022quantum, itani2022analysis, itani2024quantum, sanavio2023quantum, sanavio2024lattice,succi1993lattice,succi2015quantum} have recently attracted significant interest. The most recent quantum LGA methods aim to achieve a quantum advantage using $\log(N)$ qubits for representing $N$ cells, carrying out $O(T\log N)$ operations where $T$ is the number of time steps. Such an algorithm is usually addressed as \textit{multi-time-step} algorithm. Until the discovery of an efficient tomography technique, such an algorithm could be conceived in all the applications that do not need the knowledge of the whole quantum state, such as measurements of global properties \cite{huang2020predicting}. This advantageous algorithm, at the current state of the art, is hindered by the necessity of measurement and reinitialization in QLBM or by some fundamental properties for QLGCA \cite{schalkers2024importance, fonio2023quantum}. Thus, we are going to explore novel possibilities for QLGA. 

In this sense, it is worth developing new classical methods capable of simulating non-linearities, but with features that admit an advantageous quantum implementation. We refer these classical methods to as \textit{quantum-friendly methods}. Thus, we are interested in the linearity/unitarity of the collision operator in LGCA, but also in the simulation advantages originating from the equilibrium update rules in LBM. 
Considering this, the key to achieve a quantum advantage could be found in a model that is between LGCA and LBM. If we look at the history of LGA, the gap between these two methods was fulfilled by \textit{Integer Lattice Gas Algorithms} (ILGA) in which the exclusion principle of the LGCA is no longer applied \cite{boghosian1997integer, chopard1998multiparticle, masselot1998multiparticle}. 
The most recent works \cite{blommel2018integer, wagner2016fluctuating} achieved LBM equilibrium distributions, Poisson fluctuations and over-relaxation \cite{strand2022overrelaxation} with Monte Carlo lattice gas automata (MCLGA). It is promising, then, to start from MCLGA to design a \textit{quantum friendly} classical model, modifying the features that hinder quantum advantage. 

In this regard, the random extractions used in MCLGA have no clear counterparts in advantageous quantum algorithmic procedures for QLGA. Thus, instead of these, we carry out a fixed or equilibrium-adapted fraction of the possible collisions, retrieving this way the non-linearities. 
This, together with a novel quantum encoding of the lattice, allows us to linearize the collision operator simplifying the quantum collision process. We will show that our model recovers the same equilibrium distribution functions and the same simulation results as LBM in the $|u|\approx 0$ limit, where $u$ is the momentum density. 
The effectiveness of our model to reproduce LBM behavior is demonstrated with the simulation of a cosine-wave, that in the quantum version uses $\log(N)+3$ qubits for representing $N$ cells. The complexity of one evolution step is the complexity of the streaming step adopted \cite{shakeel2020efficient} $O(\log^2N)$, while the collision is a linear combination of unitaries (LCU), giving a cost of $O(1)$ for the 1D case studied here. Thus, the one-time-step complexity scales as $O(\log^2N)$, showing an exponential speed-up compared to the classical case. Considering the total evolution, the current implementation carries out measurements and reinitialization after each time step, increasing the total complexity by the cost of tomography and reinitialization. Initialization and tomography are open problems on their own, thus we will not deal with them in detail here, assuming for the simulations a perfect tomography, allowing for a correct reinitialization.

The paper is structured as follows: in Sec.\ref{sec:lga intro} we briefly introduce LGA, with emphasis put on the procedures from LGCA, ILGA and LBM used in the proposed method; in Sec.\ref{sec:classical ILGA} we formulate the novel standard and adaptive ILGA; in Sec.\ref{sec: quantum} we develop the QC encoding and algorithm for simulating the proposed ILGA; in Sec.\ref{sec:results} we display the numerical results of the simulations, validating the expected equilibrium distribution functions and carrying out the simulation of a cosine-wave as the non-linear benchmark case.

\section{Classical lattice gas algorithms} \label{sec:lga intro}
\subsection{Lattice gas cellular automata}
\begin{figure}[ht]
    \centering
    \scalebox{0.85}{
    \begin{tikzpicture}
        \def\l{0.5}
        
        \newcommand{\drawlattice}{
        \foreach \j in {0,1.5,3}{
        \foreach \i in {0,1,2,3,4,5}
                {
                \draw[fill=white] (\i, \j) circle (3pt);
                }
            }
        }
        
        \newcommand{\drawarrows}[3]{
        \foreach \ang in {#3}{
        \draw[-stealth, black, thick] 
        ({#1 + 0.1 * cos(\ang)}, #2) -- ({#1 + \l*cos(\ang)},#2);}}
        \newcommand{\drawrest}[2]{
        \draw[fill=black] (#1, #2) circle (3pt);
        }
        
        \drawlattice
        \drawarrows{0}{3}{0,180}
        \drawarrows{1}{3}{0}
        \drawrest{2}{3}
        \drawarrows{3}{3}{180}
        \drawarrows{4}{3}{0}
        \drawrest{5}{3}
        \drawrest{4}{3}
        \filldraw[black] (0, 2.5) node[anchor=center]{$[101]$};
        \filldraw[black] (1, 2.5) node[anchor=center]{$[001]$};
        \filldraw[black] (2, 2.5) node[anchor=center]{$[010]$};
        \filldraw[black] (3, 2.5) node[anchor=center]{$[100]$};
        \filldraw[black] (4, 2.5) node[anchor=center]{$[011]$};
        \filldraw[black] (5, 2.5) node[anchor=center]{$[010]$};
        \draw[gray, thick] (-3.5,2) -- (5.5,2);
        \drawrest{0}{1.5}
        \drawarrows{1}{1.5}{0}
        \drawarrows{2}{1.5}{0,180}
        \drawarrows{3}{1.5}{180}
        \drawarrows{4}{1.5}{0}
        \drawarrows{5}{1.5}{0,180}
        \drawrest{4}{1.5}
        \filldraw[black] (0, 1) node[anchor=center]{$[010]$};
        \filldraw[black] (1, 1) node[anchor=center]{$[001]$};
        \filldraw[black] (2, 1) node[anchor=center]{$[101]$};
        \filldraw[black] (3, 1) node[anchor=center]{$[100]$};
        \filldraw[black] (4, 1) node[anchor=center]{$[011]$};
        \filldraw[black] (5, 1) node[anchor=center]{$[101]$};
        \draw[gray, thick] (-3.5,0.5) -- (5.5,0.5);
        \drawrest{0}{0}
        \drawarrows{0}{0}{0}
        \drawarrows{1}{0}{180}
        \drawarrows{2}{0}{0,180}
        \drawarrows{3}{0}{0}
        \drawarrows{4}{0}{180}
        \drawarrows{5}{0}{0}
        \drawrest{4}{0}
        \filldraw[black] (0, -0.5) node[anchor=center]{$[011]$};
        \filldraw[black] (1, -0.5) node[anchor=center]{$[100]$};
        \filldraw[black] (2, -0.5) node[anchor=center]{$[101]$};
        \filldraw[black] (3, -0.5) node[anchor=center]{$[001]$};
        \filldraw[black] (4, -0.5) node[anchor=center]{$[110]$};
        \filldraw[black] (5, -0.5) node[anchor=center]{$[001]$};

        \filldraw[black] (-2, 4) node[anchor=center]{$x$};
        \filldraw[black] (0, 4) node[anchor=center]{$0$};
        \filldraw[black] (1, 4) node[anchor=center]{$1$};
        \filldraw[black] (2, 4) node[anchor=center]{$2$};
        \filldraw[black] (3, 4) node[anchor=center]{$3$};
        \filldraw[black] (4, 4) node[anchor=center]{$4$};
        \filldraw[black] (5, 4) node[anchor=center]{$5$};
        
        \filldraw[black] (-2, 2.75) node[anchor=center]{Before collision};
        \filldraw[black] (-2, 1.25) node[anchor=center]{After collision};
        \filldraw[black] (-2, -0.25) node[anchor=center]{After streaming};
    \end{tikzpicture}
    }
    \caption{Lattice gas Cellular Automata example of evolution for 1-dimensional model with 3 velocities named D1Q3. Each cell consists of 3 bits $[b_2b_1b_0]$, representing the presence of particles with corresponding velocities $[-1,0,1]$. The collision that preserves mass and momentum (considering a rest particle with mass 2) is $[101]\leftrightarrow[010]$, taking place at $x=0,2,5$. All the other cells are not affected by the collision. The streaming takes place according to respective velocities with periodic boundary conditions.}
    \label{fig:D1Q3_lattice}
\end{figure}
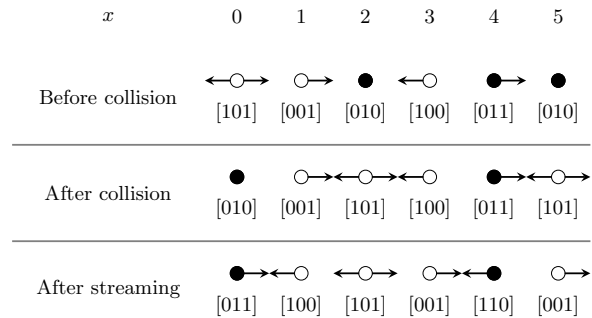

In LGCA we have a lattice where each site (or \textit{cell}), is populated by particles. These particles can have a finite set of $v$ possible velocities, giving conventionally the name to \dnqv models where $n$ is the number of dimensions. It includes an \textit{exclusion principle}, stating that there can be a maximum of one particle per site per velocity. Thus, each cell is represented by a bit string $[b_0,b_1,\dots,b_v]$ where $b_i$ is associated to the presence (1) or absence (0) of a particle with velocity $v_i$. The evolution is given by a \textit{collision} step when particles at the same site interact, and a \textit{streaming} step, when particles diffuse according to their velocity. An example is represented in Fig.\ref{fig:D1Q3_lattice}

According to this evolution, we can write an update rule for the \textit{mean occupation numbers} $B_i(x,t)=\langle b_i\rangle$ calculated on neighbors of $x$, that is

\begin{equation} \label{eq:lgca evolution}
    B_i(x+c_i\Delta t, t+\Delta t) = B_i(x,t) + \Delta_i,
\end{equation}
where $\Delta_i$ is the collision term. An H-theorem guarantees the thermalization of the system \cite{Rivet_Boon_2001} and the exclusion principle guarantees that the equilibrium distributions $B_i^{eq}$ are Fermi-Dirac. The collision usually conserves some quantities, such as mass and momentum. Based on these conservations, it is possible to derive the macroscopic dynamics of the system with a Chapman-Enskog expansion, that we use and detail in the Appendix. LGCA were successful initially, but suffered from several problems for hydrodynamic simulations. Moreover, it is easily parallelizable, but when it comes to an advantageous translation for quantum algorithms, there are fundamental problems concerning the up-to-now proposed encodings, and these limitations were investigated in \cite{fonio2023quantum,schalkers2024importance}. Thus, we need to move forward from LGCA for more accurate simulations and an analogous advantageous quantum algorithm.

\subsection{Integer lattice gas automata} \label{subsec:ilga}

If we no longer apply the exclusion principle of LGCA, each cell is represented by a set of integers $[n_0,n_1,\dots,n_v]$ where $n_i$ is the number of particles with velocity $v_i$. Thus, the evolution rule will be
\begin{equation} \label{eq:evolution ilga}
    n_i(x+c_i\Delta t, t+\Delta t) = n_i(x,t) + \Xi_i.
\end{equation}
Even if it has the same form of Eq.\ref{eq:lgca evolution}, we must notice that they are inherently different, since we move from Booleans to integers. Whereas in LGCA the collision is Boolean, thus it either takes place or it does not, in ILGA we need to define \textit{how many} collisions per site occur. One interesting possibility is given by Monte Carlo
lattice gas automata (MCLGA) \cite{wagner2016fluctuating}. Let us consider a D1Q3 MCLGA with 3 velocities $[-1,0,+1]$. In each cell, we randomly choose 2 particles. If they can collide according to the conservation of mass and momentum, they do it with a fixed probability $\lambda$. This process takes place several times in each cell, and is followed by the streaming step. Considering the random extraction, the probability of collision has a non-linear dependency on the distributions, this being the non-linear contribution in the following collision term $\Xi$ of this model

\begin{equation} \label{eq:collision matrix mclga}
    \Xi^{MCLGA} = \frac{\lambda}{\rho^2} \begin{pmatrix}
        \frac{f_0^2}{8}-2f_-f_+ \\
        -\frac{f_0^2}{4}+4f_-f_+ \\
        \frac{f_0^2}{8}-2f_-f_+
    \end{pmatrix}.
\end{equation}

Here $\rho$ is the mass density and $f_i=\langle n_i \rangle$ are the ensemble-averaged variables. Given the collision term we can derive the equilibrium distributions. This method will be applied also in further sections. For calculating the equilibrium distributions, we start going from distribution space $\Vec{f} = [f_-,f_0,f_+]^T$ to momentum space $\Vec{m} = [\rho,\rho u,\pi]^T$, where $\rho$ is the mass density, $\rho u$ is the momentum density and $\pi$ is the analog of an energy. To go from distribution space to momentum space, we use the matrix $M$ defined as

\begin{equation}
    M = 
    \begin{pmatrix}
        1 & 1 & 1 \\
        -\sqrt{3} & 0 & \sqrt{3} \\
        \sqrt{2} & -\frac{1}{\sqrt{2}} & \sqrt{2}
    \end{pmatrix}.
\end{equation}
Thus we have $M\Vec{f}=\Vec{m}$. If we multiply both sides of ensemble-averaged Eq.\ref{eq:evolution ilga} by $M$, we obtain the evolution of the momenta. The equilibrium condition is given by the relaxation of the non-conserved momentum $\pi$. Solving $M\Xi^{MCLGA}=0$ we obtain

\begin{equation}
    \pi^{eq}=\sqrt{\rho}(-1\pm\sqrt{1+3u^2}).
\end{equation}
Then, we can get the equilibrium distributions as $\Vec{f}^{eq}=M^{-1}\vec{m}^{eq}$. In MCLGA we have 
\begin{equation} \label{eq: eqdistr_mclga}
    f_i^{eq} = \rho w_i [1+3c_iu+(3c_i^2-1)(\sqrt{1+3u^2}-1)],
\end{equation}
where $w_i$ are the LBM weights given by matching the discrete velocity moments to the continuum velocity moments. We remark that is the presence of non-linear terms in Eq.\ref{eq:collision matrix mclga}
to bring non-linear equilibrium distributions that can approximately reproduce LBM considering $|u|\approx0$. This is particularly important since the macroscopic dynamic that can be simulated is based on the equilibrium distributions. Thus, if a method reproduces the same equilibrium distributions of LBM it should simulate the same phenomena.

This result was obtained due to a stochastic process and thus due to the fact that ensemble averaging of the probabilistic collisions gives a nonlinear term in the equilibrium distributions. Further improvements were reached considering a faster sampling operator \cite{seekins2022integer}, and overrelaxation was proven possible with this method \cite{strand2022overrelaxation}. However, any stochastic process is not feasible with a sub-linear quantum encoding of the space. Thus, we seek a method for obtaining LBM behavior without using random processes.

\subsection{Lattice Boltzmann method}

LBM evolves mesoscopic probability distribution functions (PDF) $g_i$ directly according to the lattice Boltzmann equation

\begin{equation}
    g_i(x+c_i \Delta t, t+\Delta t) = g_i(x,t) + \Omega_i(x,t),
\end{equation}
where $\Omega_i$ is the collision term. Commonly the single time BGK approximation \cite{bhatnagar1954model} is applied, giving
\begin{equation}
    \Omega_i(x,t) = \frac{1}{\tau}[g_i^{eq}(x,t)-g_i(x,t)],
\end{equation}
where $\tau$ is the relaxation time. It is possible to have different relaxation times for different momenta \cite{d2002multiple}, naming the \textit{multiple relaxation time} (MRT) method. Since for a classical gas we expect Maxwell-Boltzmann distributions, the equilibrium functions are given by an expansion of this latter to the desired order of the momentum, thus 

\begin{equation}\label{eq:lbm ed}
    g_i^{eq} = \rho w_i (1 + v_i u + \frac{1}{2} u^2 + o(u^3)). 
\end{equation}
For D1Q3 we have $w_0=2/3$ and $w_-=w_+=1/3$, in order to match the discrete and the continuous momenta. These are the equilibrium distributions that we aim to find with a new method that has advantages in its quantum implementation.

\section{ \label{sec:classical ILGA} Adaptive lattice gas algorithm}

The LGA introduced in Sec.\ref{sec:lga intro} shows different problems for inspiring multi-time-step quantum algorithms. For this reason, we design in this section an algorithm that is able to reproduce LBM equilibrium distributions but uses a linear operator in its quantum translation. The best candidate, that we prove successful for this purpose, is then an ILGA that, differently from MCLGA, does not involve random extractions. With the results of this section, we are going to define in the next one the analogous quantum algorithm.

We consider an ILGA where each cell has an integer number of particles $[n_-, n_0, n_+] \in \mathbb{N}^3$ with corresponding lattice velocities $[-1,0,+1]$. We consider two collisions: \textit{crunches} of opposite-moving particles into rest particles, and \textit{splitting} of two rest particles into a couple of opposite-moving particles. For simplicity, we are going to consider the rest particles having mass 1, as in \cite{wagner2016fluctuating}, and being always even. The maximum number of splittings that can take place is $n_0/2$; the maximum number of crunches that can take place is $\text{min}(n_+,n_-) = \frac{1}{2}(n_++n_--|\tilde{u}|)$ being $\tilde{u} = n_+ - n_-=\rho u/\sqrt{3}$ the local momentum. We consider that only a fraction $\lambda_s \in \mathbb{R}$ of splittings takes place, and only a fraction $\lambda_c \in \mathbb{R}$ of crunches takes place. This is a straightforward way of deciding upon the collisions for an ILGA: we consider fractions of the maximum number of collisions that can take place. The calculation of this minimum is the non-linear part of the algorithm. Thus, we can write the collision term as follows
\begin{equation} \label{eq:collision term matrix}
    \Xi = \begin{pmatrix}
        \lambda_s \frac{n_0}{2} - \lambda_c \text{min}(n_+,n_-) \\
        -\lambda_s n_0 + 2\lambda_c \text{min}(n_+,n_-) \\
        \lambda_s \frac{n_0}{2} - \lambda_c \text{min}(n_+,n_-)
    \end{pmatrix}.
\end{equation}
We can also write it in compact form
\begin{equation}
    \Xi_i = (3c_i^2 - 2) \qty[\lambda_s \frac{n_0}{2} - \lambda_c (n_++n_--|\tilde{u}|)]
\end{equation}

for $i\in{0,1,2}$. We notice that in general $\Xi_i\in \mathbb{R}$, thus if $n_i$ are integers at one time step, they are not integer anymore at the next time step. Since here in the theoretical treatment, as we are considering particles, we can consider $\Xi_i$ to be cast into integers according to conservation laws and the even constraint for $n_0$. We see in Sec.\ref{sec:results} that numerically this is not strictly necessary.

From this collision term we can calculate the equilibrium distributions as reported in Sec.\ref{subsec:ilga}. We solve $M\Xi=0$ obtaining the equilibrium momentum

\begin{equation}
    \pi^{eq} = \rho \frac{\sqrt{2}}{2} \frac{(2\lambda_s - \lambda_c + 3 \lambda_c |u|)}{\lambda_c + \lambda_s}. 
\end{equation}
From here we obtain the equilibrium distributions of our ILGA
\begin{equation} \label{eq:feq_0}
    f^{eq} = \begin{pmatrix}
        -\frac{\rho u}{2} + \frac{\rho}{6} + \frac{\sqrt{2}}{6}\pi^{eq} \\
        \frac{2}{3} \rho - \frac{\sqrt{2}}{3}\pi^{eq} \\
        \frac{\rho u}{2} + \frac{\rho}{6} + \frac{\sqrt{2}}{6}\pi^{eq}
    \end{pmatrix}.
\end{equation}
We notice that these equilibrium functions are \textit{linear} in $u$, while we aim to find quadratic equilibrium distribution functions as in LBM. However, let us study the equations that this ILGA follows further. At the order $o(\epsilon \Delta t)$ with the Chapman-Enskog expansion, detailed in the Appendix, we obtain
\begin{align}
    \pdv{\rho}{t} + \pdv{\rho u}{x} & = 0 ,\label{eq:mass conservation}\\
    \pdv{\rho u}{t} + \frac{\lambda_s}{\lambda_c + \lambda_s} \pdv{\rho}{x} + \frac{\lambda_c}{\lambda_c + \lambda_s} \pdv{|\rho u|}{x} & = 0 .\label{eq:mom conservation}
\end{align}
Eq.\ref{eq:mass conservation} comes from mass conservation, while Eq.\ref{eq:mom conservation} comes from momentum conservation. The first one is the one expected, a standard continuity equation. The second one provides the second term that is analogous to a pressure term, and the third term that is supposed to be analogous to a convective term. However, in that case the term $\lambda_c/(\lambda_c+\lambda_s)$ should be proportional to $u$, while $\lambda_c$ and $\lambda_s$ are constant. For this fact, and for not getting quadratic equilibrium distributions, we apply another strategy.

\subsection{Local-LBM adaptation}
The equilibrium condition can be obtained by equaling Eq.\ref{eq:collision term matrix} to 0, resulting in
\begin{equation} \label{eq: equilibrium condition}
    \lambda_s \frac{n_0^{eq}}{2} = \lambda_c \text{min}(n_+,n_-)^{eq}.
\end{equation}
We can infer from Eq.\ref{eq: equilibrium condition} that the fractions $\lambda_s$ and $\lambda_c$ govern the ratios among equilibrium distributions. Earlier in this section, we fixed $\lambda_s$ and $\lambda_c$ obtaining linear equilibrium distributions, but with the presence of $|u|$. The idea here is to reverse this process: we fix the equilibrium distributions and modify the collision parameters accordingly. This adaptation is made by considering that for simulating CFD we expect the local equilibrium distribution to be Maxwell-Boltzmann (MB). In LBM this is the motivation for using MB distribution for the collision with the BGK approximation. In the same sense, the populations in our algorithm collide according to the desired equilibrium distributions. It is not given, however, that \textit{imposing} the equilibrium distributions would lead to the same equilibrium distribution. We substitute $n_i^{eq}$ from Eq.\ref{eq: equilibrium condition} with $g_i^{eq}$ in Eq.\ref{eq:lbm ed}, and we obtain

\begin{equation} \label{eq:lc_lbm}
    \lambda_c = \lambda_s \frac{2-3u^2}{1 - 3|u| + 3u^2}.
\end{equation}
This is obtained because, depending on the sign of $u$, we have either $n_+^{eq}$ to consider ($u<0$), or $n_-^{eq}$ ($u>0$). Before thermalization, these values change from point to point, resulting in a \textit{local adaptation}. We prove numerically in Sec.\ref{sec:results} that this adaptation replicates LBM behavior in approximately the same velocity range as MCLGA. We notice that for having a physical meaning, we need $0<\lambda_c<1$. In general this is ensured if $|u|<\sqrt{\frac{2}{3}}$. However, we can have sites where $|u|>\sqrt{\frac{2}{3}}$, causing $\lambda_c$ to be negative or bigger than one. In this case, the collision does not take place. Thus, in the region far from $|u|\approx 0$, as we will see in the Results section, Eq.\ref{eq:lc_lbm} does not hold anymore, causing a difference with LBM.

This approach of \textit{fixing} the equilibrium distributions turns out successful. This means that the proposed algorithm with local adaptation behaves as LBM, simulating the same equations, and inheriting the same limitations for flow velocities. Now, we will see the quantum version of this algorithm.

\section{\label{sec: quantum}Quantum implementation}

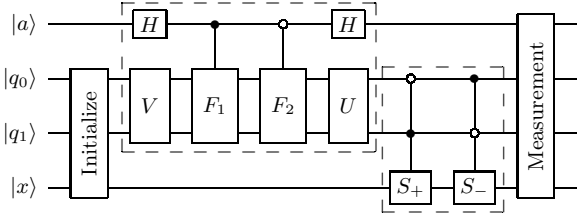
\begin{figure}
    \centering
    \scalebox{0.9}{
    \Qcircuit @C=1.0em @R=1.0em @!R {
        & & & & & & & &\\
        \lstick{\ket{a}} & \qw & \gate{H} & \ctrl{1} & \ctrlo{1} & \gate{H} & \qw & \qw & \multigate{3}{\rotatebox{90}{Measurement}} & \qw \\
        \lstick{\ket{q_0}} &  \multigate{2}{\rotatebox{90}{Initialize}} & \multigate{1}{V} & \multigate{1}{F_1} & \multigate{1}{F_2} & \multigate{1}{U} & \ctrlo{1} & \ctrl{1} & \ghost{\rotatebox{90}{Measurement}} & \qw \\
        \lstick{\ket{q_1}} & \ghost{\rotatebox{90}{Initialize}} & \ghost{V} & \ghost{F_1} & \ghost{F_2} & \ghost{U} & \ctrl{1} & \ctrlo{1} & \ghost{\rotatebox{90}{Measurement}} & \qw \\
        \lstick{\ket{x}} & \ghost{\rotatebox{90}{Initialize}} & \qw & \qw & \qw & \qw  & \gate{S_+} & \gate{S_-} & \ghost{\rotatebox{90}{Measurement}} & \qw
        \gategroup{2}{3}{4}{6}{.7em}{--}
        \gategroup{3}{7}{5}{8}{.7em}{--}
    }}
    \caption{Quantum circuit for executing the algorithm. After initialization, collision is carried out with a LCU, thus the operator $\hat{C}$ in Eq.\ref{eq:collision operator} is decomposed as $C=UFV$ where $F=F_1+F_2$, being the whole gate including the ancilla reported as "collision\_gate" in Alg.\ref{alg:qalga}. Then the streaming is carried out with \cite{shakeel2020efficient} method, having $S_+$ for the streaming to the right and $S_-$ for the streaming to the left being the whole gate including the controls reported as "streaming\_gate" in Alg.\ref{alg:qalga}. At the end of the circuit, we have the measurement of the lattice state and the re-initialization for the successive time step}
    \label{fig:qcircuit}
\end{figure}

In this section we show how the presented algorithm can be translated into a quantum algorithm that needs $O(\log N)$ qubits where $N$ is the number of sites in the lattice.
The first step in developing this quantum algorithm is the encoding of the lattice. In \cite{budinski2021quantum}, the probability distributions are directly encoded in the quantum amplitudes of a state. Equivalently, we encode the populations of our LGA in quantum amplitudes as follows
\begin{eqnarray} \label{eq:encoding}
\ket{\Psi} = \frac{1}{M} && \sum_x  \ket{x}(n_0(x) \ket{00} + n_1(x) \ket{01} \nonumber\\
     && + n_2(x) \ket{10} + \lambda_c \text{min}(n_0(x),n_1(x)) \ket{11}),
\end{eqnarray}
where $M=\sqrt{\sum_{x,i}n_i(x)^2}$.
Unlike the state-of-the-art QLBM schemes, we include the non-linear term, thus the \textit{min} term, in the amplitude of $\ket{11}$ state: the population of the rest particles is encoded in $\ket{00}$ amplitude, the right moving particles in $\ket{01}$ amplitude, the left moving particles in $\ket{10}$ amplitude, and the non-linear term in $\ket{11}$ amplitude. The latter is calculated with the corresponding $\lambda_c$, depending on whether local adaptation is applied. 
The resulting algorithm provides, as \cite{budinski2021quantum}, a succession of (re)initialization, collision, streaming and measurement for each time step as depicted in Fig.\ref{fig:qcircuit}. 
This algorithm can achieve time-step concatenation, avoiding measurement and reinitialization, when the amplitude of the state $\ket{11}$ can be calculated unitarily. When $\lambda_c$ is constant (not in the local adaptation case), the condition for having time-step concatenation is $\text{min}(n_0,n_1)=n_0$ or $\text{min}(n_0,n_1)=n_1$ in any cell at any time. If the adaptation is local, $\lambda_c$ depends on the local momentum. In this case, if the local momentum is known analytically, without need to be measured each time step, and if it has the same sign everywhere as in the standard case, then the algorihtm can be time-step concatenated. We leave the study of these cases as a perspective for future research.

The collision operator using the encoding in Eq.\ref{eq:encoding} is

\begin{equation} \label{eq:collision operator}
    \hat{C} = \begin{pmatrix}
        1- \lambda_s & 0 & 0 & 2 \\
        \lambda_s/2 & 1 & 0 & -1 \\
        \lambda_s/2 & 0 & 1 & -1 \\
        0 & 0 & 0 & 1
    \end{pmatrix}.
\end{equation}

This operator is not unitary. To implement it on a quantum circuit we apply the singular-value decomposition (SVD) \cite{nielsen2010quantum}, obtaining $C=UFV$ where $U$ and $V$ are unitary, and $F$ is diagonal, but not unitary. Then, $F$ is implemented as a linear combination of unitaries (LCU) \cite{childs2012hamiltonian} using an ancilla. The streaming step is executed with the following operator
\begin{equation}
    \hat{S} = \sum_x \ket{x+1}\bra{x} \otimes \ket{01}\bra{01} + \ket{x-1}\bra{x} \otimes \ket{10}\bra{10},
\end{equation}
that can be efficiently carried out using the shift algorithm described in \cite{shakeel2020efficient}. This shift operation, however, does not transport the non-linear contribution of the $\ket{11}$ amplitude. This non-linear term must be recalculated following the streaming step. If a quantum algorithm could compute this non-linear term unitarily, or if this term was constant in time, reinitialization would become unnecessary. In the current implementation, we opt for measurement and subsequent reinitialization of the lattice state after the combined collision and streaming operations. The pseudocode of the quantum algorithm is provided in Alg.\ref{alg:qalga}.

\begin{algorithm}[H]
\caption{Quantum Adaptive LGA}\label{alg:qalga}
\begin{algorithmic}
\State $N,T$
\State $\texttt{c\_lattice}$
\State $\texttt{initialize(c\_lattice)}$
\State $\lambda_s,\lambda_c$
\State $\hat{C}(\lambda_s)$ \Comment{from Eq.\ref{eq:collision operator}}
\State $\hat{F},\hat{U},\hat{V} \gets \text{SVD}(\hat{C})$ \Comment{from \cite{nielsen2010quantum}}
\State $\hat{F_1},\hat{F}_2 \gets \texttt{LCU}(\hat{F})$ \Comment{from \cite{childs2012hamiltonian}}
\State $\texttt{collision\_gate} \gets \hat{F_1},\hat{F_2},\hat{U},\hat{V}$
\State $\hat{S}_+ , \hat{S}_-$  \Comment{from \cite{shakeel2020efficient}}
\State $\texttt{streaming\_gate} \gets \hat{S}_+ , \hat{S}_-$
\State $\texttt{qc}\gets\texttt{QuantumCircuit(q\_reg,x\_reg,a\_reg)}$
\State $\texttt{qc.append(collision\_gate)}$
\State $\texttt{qc.append(streaming\_gate)}$

\While{$t < T$}
\State $\texttt{q\_state} \gets \texttt{encoding(c\_lattice)}$ \Comment{from Eq.\ref{eq:encoding}, $O(N)$}
\State $\texttt{qstate.evolve(qc)}$ \Comment{$O(\log^2N)$}
\State \texttt{classicallattice} $\gets$ \texttt{tomography(qstate)} \Comment{$\Omega(N)$}
\EndWhile
\end{algorithmic}
\end{algorithm}

\subsection{Computational complexity}  

In this subsection we analyse the space and computational complexity of ALGA and QALGA. For the classical case, the space resources scale linearly with the number of lattice sites $N$, while the computational complexity scales as $O(NT)$, where $T$ is the number of time steps. The individual time step has a cost of $O(N)$ since the collision must be carried out in each site and streaming has also a linear dependence on $N$.

The quantum algorithm, on the other side, uses $\log N +3$ qubits, considering also the ancilla, thus allowing an advantageous scaling in terms of qubits respect to the classical encoding of the lattice. 

Concerning computational complexity, we analyze the cost of collision, streaming, and the whole evolution step including measurement and reinitialization. The quantum collision has a cost of $O(1)$, being used only two qubit gates since we deal with only two occupation qubits $\ket{q_0q_1}$. Qiskit default decomposition for the collision operator counted from 120 to 124 operations for different $\lambda_s$. For the streaming step, there are different options for implementing the quantum walk procedure. Most recent proposals \cite{shakeel2020efficient,razzoli2024efficient,budinski2023efficient} provide a cost $O(\log^2N)$. Different trade-offs can be found with more qubits, as in \cite{budinski2023efficient} lowering the cost to $O(\log N)$. Since we adopt the method proposed in \cite{shakeel2020efficient} we consider the total cost of the evolution step to be $O(\log^2N)$, thus representing an exponential speed up respect to the classical algorithm. 

For evaluating the complexity of the whole algorithm is necessary to include measurement and reinitialization for the computation of the non-linear term in the encoding. Full quantum state tomography scales exponentially with the number of qubits \cite{paris2009quantum}, thus linearly with the number of sites. In particular, there are different techniques requiring the measurement of $2^{2\log N}$ observables \cite{james2001measurement}, bringing to a linear cost in the number of sites. Improvements to this cost can be done for specific cases \cite{xin2017quantum,binosi2024tailor}, but we found no case specifically applicable to our algorithm. A variety of techniques can possibly lower this cost \cite{gross2010quantum, aaronson2018shadow}. Their direct implementation for this quantum algorithm is left as a perspective for future research. Initialization, on the other side, also scales exponentially with the number of qubits \cite{nielsen2010quantum}. Some trade-offs introducing ancillae were proposed \cite{zhang2024circuit}, lowering the computational cost of initializing an arbitrary quantum state but increasing considerably the number of qubits needed. Direct applications of these techniques are of general interest for amplitude-encoding algorithms as ours, and thus a promising perspective. In the end, the necessity of measurement and reinitialization still forbids a real advantage for the current implementation, this being, however, an active area of research.

\section{Numerical results} \label{sec:results}

In this section, we show the numerical results of the model we develop. Before doing so, we remark on an important aspect of the following simulations. When we calculate the number of colliding particles, we multiply an integer ($n_0$ for splits and $\min (n_0,n_1)$ for crunches) by a float ( $\lambda_s$ or $\lambda_c$), obtaining a float. This float should be cast into an integer to represent particles according to conservation laws and for maintaining $n_0$ even. In this case, we obtain noisy data that need to be averaged over different realizations. Since the model is deterministic, the only difference we have in different realizations lies in the initial conditions. Averaging over different initial conditions is usually used in LBM for turbulent flows or optimization, which we do not consider here and leave as perspectives for future works. Thus, in our case, different realizations are deterministic evolutions of random initial conditions that try to approximate the same initial condition. Moreover, we can keep in mind that we aim to encode the local populations in the amplitudes of a quantum system, which are complex in general. 

Considering these aspects, we carried out the simulations with and without integer casting. When we average over different realizations with integer casting and random initialization, as expected for the reasons above, we obtain the same results as without integer casting with only one realization and smooth initial distribution functions, having a clear advantage in computing time. This holds also considering that the calculations we carried out for the equilibrium distributions in Sec.\ref{sec:classical ILGA} did not involve the integer constraint in any point.

The binary or integer characteristics of historical LGCA and ILGA were given for simulation purposes for the available technology, and were strictly linked to the non-linearity observed. However, in our case the non-linearity of the model lies in the collision process, making it more convenient for simulations to drop the integer constraint. Nonetheless, we should keep in mind that designing this algorithm requires thinking in terms of particles, thus integer numbers, and that this constraint was removed only for simulation purposes. For other simulations, it could be necessary to maintain the integer casting. Considering these aspects, the following results do not consider integer casting and do not average over random initializations. 

\subsection{Constant parameters} \label{subsec:sim_1}
\begin{figure}[ht]
    \centering
    \includegraphics[width=0.9\linewidth]{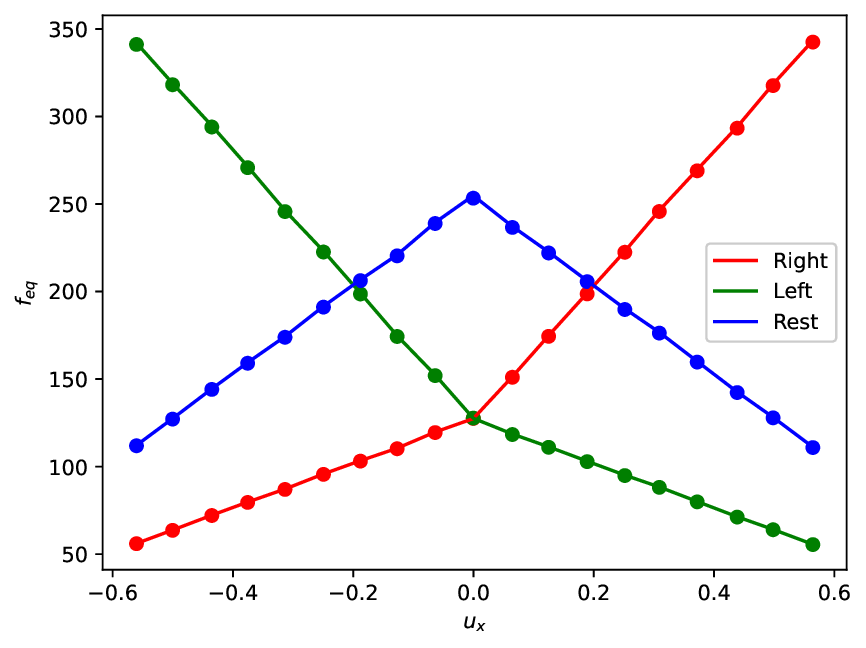}
    \caption{Equilibrium distribution functions averaged over time for D1Q3 ILGA model with fixed parameters $\lambda_c = \lambda_s = 0.2$ . The full line is theoretical expectations from Eq. \ref{eq:feq_0}, and points are the simulations' results. The x-axis coordinate is $u_x=\frac{\sum_x n_+(x)-n_-(x)}{\sum_x\sum_i n_i(x)}$. In this simulation, the number of sites is 200, $n_{max}=500$ and we let the simulation evolve for 2000 time steps, starting the average from $t=1600$.}
    \label{fig:feq_lin}
\end{figure}

The first validation we provide regards the equilibrium distributions in Eq.\ref{eq:feq_0} considering constant collision parameters $\lambda_c$ and $\lambda_s$. We aim to observe the system after thermalization, validating Eq.\ref{eq:feq_0}. This requires to consider the thermalization for different momentum densities, as done in \cite{wagner2016fluctuating}. We simulate the evolution of a sine-wave with periodic boundary conditions, with initial populations depending on the momentum density. In particular, considering $N$ sites and $n_{max}$ initial maximum number of particles per site, the initial populations are

\begin{equation} \label{eq: initialization}
    n_i(x_j,0) = w_in_{max}p_i\sin{\frac{\pi j}{N}},
\end{equation}
where $w_i$ are the LBM weights for D1Q3, $x_j=\frac{jL}{N}$ where $L$ is the defined length, unitary in our case, and $p_i$ are used to set the desired momentum. Specifically, $p_-=\frac{1-U}{2}$, $p_0=0.2$, $p_+=\frac{1+U}{2}$ for $U=1-\frac{m}{M}$ where $m\in\{0,\dots,M\}$, and $M$ is the number of different initial momentum densities (number of points in Fig.\ref{fig:feq_lin}). We notice that $p_0$ does not affect the equilibrium behavior since it does not add any momentum. Different $p_0$ can give a different range of measured average momentum density since it affects the total number of particles. As we can see in Fig.\ref{fig:feq_lin}, the linear behavior of the equilibrium functions is confirmed.

\subsection{Shock waves with constant parameters} \label{subsec:sim_2}
Using constant lambdas we can show that a cosine-wave initialization, used also in further subsections, yields the formation of shock waves qualitatively similar to LBM, but clearly different from these. This is due to the absolute value dependence of the equilibrium distribution functions. Even if they are not quadratic in the momentum density, the attained equilibrium distributions could be seen as their simplified linear approximation. 

For this simulation we initialize a cosine-wave using periodic boundary conditions. The initial configuration is similar to Eq.\ref{eq: initialization}, but considering $\cos{\frac{j\pi}{N}}$ and $p_i=1 \forall i$.
\begin{figure*}[ht]
    \centering
    \includegraphics[width=0.4\linewidth]{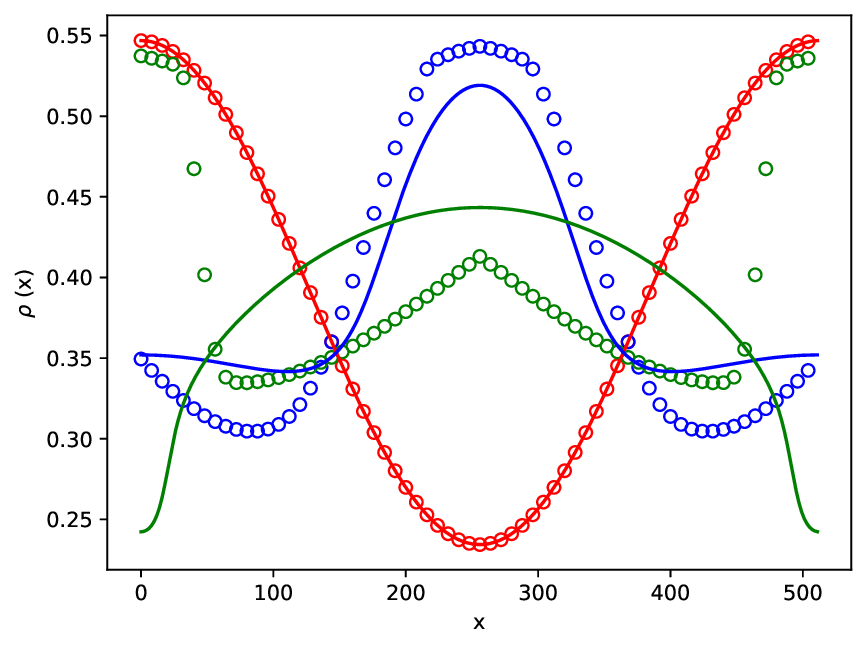}
    \includegraphics[width=0.4\linewidth]{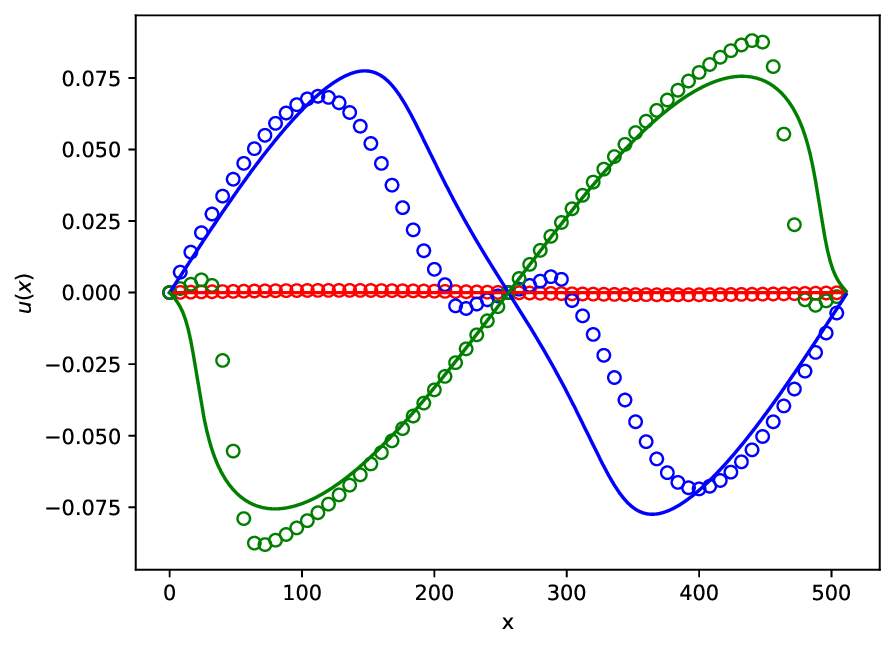}
    \caption{Comparison between Q-ALGA(circles) with constant collision parameters ($\lambda_s=0.2$, $\lambda_c=0.2$, $n_{max}=200$) and LBM (line) simulation with $\tau=1$ on 512 gridpoints for a cosine-wave initialization for mass (left) and momentum (right) densities. The quantum simulations used 12 qubits and it was made with Statevector simulator on Qiskit. The red, blue, and green line/dots are, respectively, at time $t=0,300,600$} 
    \label{fig:coswave_q_lbm_const}
\end{figure*}
We can clearly see from Fig.\ref{fig:coswave_q_lbm_const} that the constant parameters produce shock waves, but also that these are different from LBM simulations. This was expected for the difference in the equilibrium distributions. Despite the difference, this model proves to be able to capture non-linearities. A detailed analysis of the viscosity would need the solution to Eq.\ref{eq:chapman_2_1}, which is left as a future perspective. As a qualitative consideration, we can observe that the damping of the wave for some values of $\lambda_s$ and $\lambda_c$ is comparable to the wave simulated by LBM. However, since the equations are different, this does not imply a similar viscosity.

Any LGA has a clear source of non-linearities. In our case, this is the evaluation of $\min(n_0,n_1)$, bringing in an absolute value in the resulting PDE. Even if this term does not depend on a power of $u$ bigger than one, since the degree is 1, it still captures non-linearities. Thus, we believe the presence of these non-linearities is very interesting and that they deserve to be studied in detail. However, the model is incomplete, since our principal aim is to find an ILGA model that reproduces LBM behavior for paving the way to the more complex simulations LBM is capable of. Thus, we will see now the local-adapted simulations.

\subsection{Local-adapted parameters}
The first validation about local-adapted $\lambda_c$ concerns the equilibrium distribution functions. As said in Sec.\ref{sec:classical ILGA}, we \textit{impose} the equilibrium distributions adapting the fraction of crunches according to Eq.\ref{eq:lc_lbm}.  In this way, $\lambda_c$ depends on the local momentum of the cell. According to this procedure, $\lambda_c$ can be bigger than 1 or negative, losing the meaning of fraction of possible collisions on which the algorithm is based. The general condition for the collision taking place is $|u(x,t)|<\sqrt{\frac{2}{3}}$. When $\lambda_c$ is outside its boundaries, the collisions do not take place.

\begin{figure}[ht]
    \centering
    \includegraphics[width=0.9\linewidth]{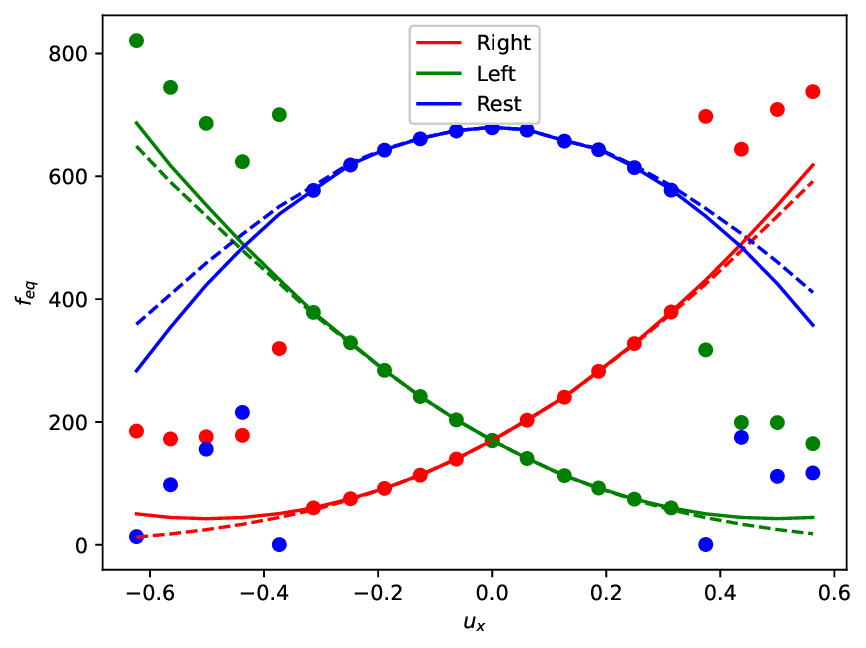}
    \caption{Equilibrium distribution functions for D1Q3 with adaptive parameters. Each point is the result of a different initialization of the lattice for having a defined velocity. The value of $\lambda_c$ is calculated according to Eq.\ref{eq:lc_lbm}, using $u(x,t)= \frac{n_+(x,t)-n_-(x,t)}{\sum_i n_i(x,t)}$ recalculated for each cell at each time step. The dashed line is theoretical Monte Carlo Lattice Gas Automata \cite{wagner2016fluctuating} 
    from Eq.\ref{eq: eqdistr_mclga}; the full line is theoretical LBM in Eq.\ref{eq:lbm ed}; points are the experiment results.}
    \label{fig:lbm local adapted}
\end{figure}

To see the equilibrium distribution functions, we carry out the same simulation as for constant parameters in Sec.\ref{subsec:sim_1}. As we can see in Fig.\ref{fig:lbm local adapted}, we obtain a good agreement with LBM equilibrium distributions for $|u|\approx0$. For the disagreement of the tails, it is possible to see that this corresponds to a majority of $\lambda_c$ being bigger than 1 or negative, forbidding the system to thermalize as expected. This happens probably because most of the collisions cannot take place, bringing to a threshold of sites where the collision does not take place, under which the state is frozen with a disordered configuration, forbidding the correct thermalization. The detailed study of this regime is out of the scope of the paper since we are interested in $|u|\approx0$ region. 

This result proves that in the region of interest $|u|\approx 0$, the proposed algorithm manages to reproduce LBM equilibrium distribution functions. This result is important since the macroscopic behavior of the system, thus the PDE governing it, depends on the equilibrium distribution functions. Consequently, this method inherits all the possibilities of LBM, respecting its limitations. This can be demonstrated with the last validation in the following subsection.

\subsection{Cosine-wave simulation}

The last validation shows that QALGA can simulate the evolution of a cosine-wave as LBM. The importance of this simulation lies in two aspects. The first is the capability of the algorithm to reproduce a non-linear behavior. In fact, we expect the formation of shock waves deriving from the collision process. The second is to reproduce the behavior of LBM, using this latter as a benchmark. 

\begin{figure}[ht]
    \centering
    \includegraphics[width=0.8\linewidth]{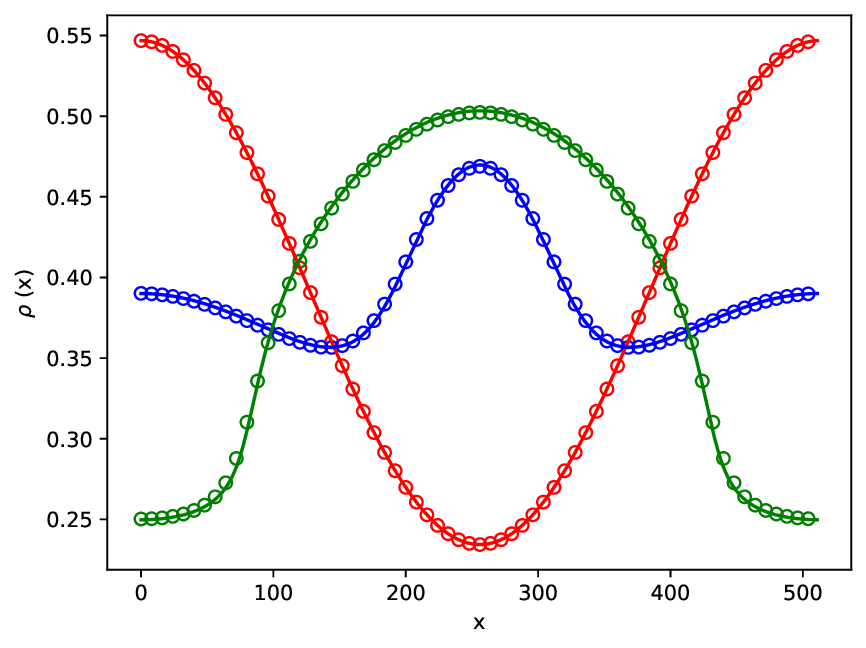}
    \includegraphics[width=0.8\linewidth]{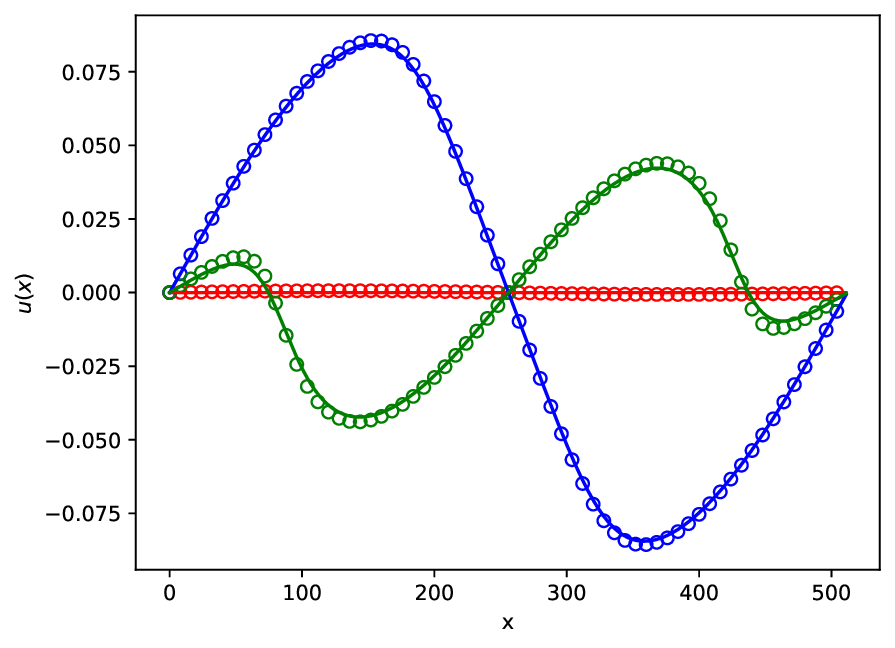}
    \caption{Comparison between QALGA(circles) with local update ($\lambda_s=0.2$, $n_{max}=200$) and LBM (line) simulation with $\tau=1$ on 512 gridpoints for a cosine-wave initialization for mass (left) and momentum (right) densities. The red, blue and green line/dots are, respectively, at time $t=0,250,500$ The quantum simulations used 12 qubits and it was made with Statevector simulator on Qiskit.} 
    \label{fig:coswave_q_lbm}
\end{figure}

For this simulation we initialize a cosine-wave with periodic boundary conditions as in Sec.\ref{subsec:sim_2}.
For the simulation in Fig.\ref{fig:coswave_q_lbm} we used 512 sites, we set $\lambda_s=0.2$ and for LBM we considered $\tau=1$. We see a good agreement between the different methods involved. It can be seen, changing the fractions of moving particles (i.e. creating a net momentum), that the agreement in the simulations persists mantaining a maximum momentum density such that $|u(x,t)|<3$ $ \forall x,t$. For higher velocities the simulation was not stable. This is consistent with the stability of LBM, thus, we can deduce that ALGA has the same limitations of LBM for simulating flows with low Mach number. The quantum algorithm was capable of reproducing the LBM behavior using a linear collision operator thanks to the encoding in Eq.\ref{eq:encoding}. We must notice that this encoding does not allow to avoid collisions when $\lambda_c$ is greater than one or negative, as we do in the classical algorithm. In this simulation we are well within the limits for which the collisions take place in each site. However, this is not guaranteed in general for every simulation.

Heuristically, it is possible to appreciate that there is a relation between $\lambda_s$ and $\tau$. Unfortunately, the formulation we derived for the moment does not allow for an analytical expression for this dependency. We could be brought to compare the viscosity coefficients at the second order of the Chapman-Enskog expansion in Eq.\ref{eq:chapman_2_1} to the viscosity of LBM $\nu=\frac{\tau-0.5}{3}$. In this case, considering local adaptation in the limit $|u|\approx0$, we would have $\nu=1/3$, resulting in $\tau=1.5$. However, the numerical analysis of the spatially averaged absolute difference between LBM and QALGA mass and momentum density did not prove any confirmation for this hypothesis.

In order to further analyze this correspondance numerically, we simulate the cosine wave for different values of $\lambda_s$ and $\tau$ for 500 steps. Then, we compare the spatially averaged absolute difference between QALGA and LBM at $t_0=450$, finding the value of $\tau$ that minimizes this distance. This showed qualitatively a possible dependence as $\tau \propto \frac{1+\lambda_s}{\lambda_s}$, resembling effectively the inverse of the expression found for constant parameters in Eq.\ref{eq:chapman_2_1}. However, a direct correspondence requires further analytical treatment, since in this case $\lambda_c$ varies, thus we leave this consideration as a qualitative basis for further studies. The QALGA simulation was stable for a maximum value of $\lambda_s^{\text{max}}=0.436 \pm 0.025$, corresponding to a $\tau^{min}=0.75\pm 0.01$, obtaining a minimal viscosity of $\nu^{min}=0.08\pm0.01$ in lattice units. For the maximum values we have $\lambda_s^{\text{min}}=0.026 \pm 0.053$ and $\tau^{max}=11.36\pm 0.01$, obtaining a viscosity of $\nu^{max}=3.62\pm0.01$ in lattice units. The best $\tau$ for each $\lambda_s$ was not constant through the whole simulation, possibly suggesting the correspondance with a multiple relaxation time LBM. Even if validations were made considering single relaxation time LBM, these are still valid for qualitatively appreciate the capacity of our algorithm to reproduce LBM results, and to give a general range for the reachable viscosities. The sensibilities of the measures in this case are given by the fact that the QALGA simulations were carried out with $\lambda_s\in[0,1]$ considering 20 equidistant values, and for LBM $\tau\in[0,20]$ considering 80 equidistant values. A possible explanation for the limitations of $\lambda_s$ can be found in the limits of $\lambda_c$, which can become unphysical due to local adaptation. These limitations deserve a detailed study for improving the model.

For the quantum simulation we used Statevector simulator on Qiskit, using $\log(N)+2$ qubits and one ancilla. For simulating different time steps, we remark that we relied on a perfect tomography of the quantum state, thus allowing a reinitialization without losing the corresponding amplitudes of any site. Eventually, the absence of this perfect tomography would lead to some missing site, resulting in mass and momentum loss. Specifically, sites with few particles have a small quantum amplitude, thus making it more difficult to measure them. If they do not get measured, that is a mass and momentum loss that affects the reinitialization. This can be improved with amplitude amplifications procedures to be left as an optimization perspective.

\section{Conclusions}

We presented a novel integer lattice gas algorithm for D1Q3 model, capable of reproducing the same equilibrium distribution functions as LBM in the $|u| \approx 0$ limit. This system of $N$ cells can be encoded in a quantum state of $\log(N)+2$ qubits, yielding a corresponding quantum algorithm, such that the collision operator is linear. However, measurement and reinitialization after each time step are still needed for the correct encoding of the classical lattice. Carrying out the collision and streaming brings to a computational cost of $O(\log^2 N)$. In cases where reinitialization and tomography are needed, the total cost of $T$ time steps is $\Omega(NT)$. It is the general perspective for all quantum algorithms using amplitude encoding and that need to retrieve the whole state, to obtain an efficient tomography. In our case, this means measuring efficiently the quantum state in order to guarantee up to a certain precision the conservation of mass and momentum from one step to the next one. This is, then, a clear direction for improvements and future research.

In this article, we studied the version of the algorithm that does not achieve time-step concatenation. However, as a perspective for future research, we highlight the potential of the algorithm to achieve such an advantage if the $\min$ term in Eq.\ref{eq:encoding} can be calculated unitarily or defined a priori by the simulation. In this case we expect to be possible to avoid measurement and reinitialization, thus bringing the cost of evolving for $T$ time steps down to $O(T\log^2(N))$, which is an important speed-up. The trade-off with the current formulation would be in the number of ancillae needed. For concatenating the collision with LCU we need 1 ancillae per time step, thus increasing the depth of the circuit to $\log N + 2 + T$ qubits. This forbids any advantage in retrieving the information needed since the probability of retrieving the desired state scales as $2^{-T}$. Different ways to make the collision unitary avoiding LCU, such as block encoding, can be considered for future research. Another limitation of this advantage is posed by the different equilibrium distributions obtained, so from the fact that the standard ALGA does not reproduce LBM equilibrium distributions.

Our results demonstrate the possibility of developing new quantum-friendly classical algorithms for achieving a quantum advantage for computational fluid dynamics. Future perspectives include (1) a detailed study of the ILGA system, its thermodynamic and its stability; (2) the possibility of avoiding reinitialization in the quantum algorithm; (3) the development of an analogous method in 2d; (4) the application of the novel encoding to QLBM schemes for a complete comparison between QALGA and QLBM.


\begin{acknowledgments}
This article is the result of a visiting research period of NF funded by Quanscient Oy, where most of the research has been carried out. A sincere acknowledgement goes to the whole research team, especially Antonio David Bastida Zamora for fruitful discussions regarding the understanding of the model and the perspectives in quantum CFD. 
\end{acknowledgments}

\appendix

\section{Chapman-Enskog expansion}
In this Appendix, we derive the PDE that govern different orders of time and Knudsen number $\epsilon$. Considering \cite{wolf2004lattice} and \cite{mohamad2011lattice} we have the following Taylor expansion

\begin{eqnarray} \label{eq:taylor}
    f_i(x&&+c_i\Delta t,t+\Delta t) - f_i(x,t) = \Delta t \qty(\pdv{f_i}{t} + c_i \pdv{f_i}{x}) + \nonumber\\
    && + \frac{\Delta t^2}{2}\qty(\pdv[2]{f_i}{t} + 2 c_i \pdv{f_i}{t}{x} + c_i^2 \pdv[2]{f_i}{x}) + o(\Delta t^3) .
\end{eqnarray}
The Chapman-Enskog expansion corresponds to

\begin{eqnarray}
    f_i & = f_i^{eq} + \epsilon f_i^{(1)} + \epsilon^2 f_i^{(2)} + o(\epsilon^3), \\
    \pdv{}{t} & \longrightarrow \epsilon \pdv{^{(1)}}{t} + \epsilon^2 \pdv{^{(2)}}{t} + o(\epsilon^3), \\
    \pdv{}{x} & \longrightarrow \epsilon \pdv{^{(1)}}{x}.
\end{eqnarray}
From here on, we substitute the Chapman-Enskog expansion in Eq.\ref{eq:taylor}, and at the order $o(\Delta t \epsilon)$ we find

\begin{equation} \label{eq:epsilont order}
    f_i(x+c_i\Delta t,t+\Delta t) - f_i(x,t) = \pdv{f_i^{eq}}{t} + c_i \pdv{f_i^{eq}}{x}.
\end{equation}
The conservation of mass states that

\begin{equation}
    \sum_i f_i(x+c_i\Delta t,t+\Delta t) - f_i(x,t) = 0.
\end{equation}
Thus we have (knowing $\rho=\sum_i f_i^{eq}$ and $\rho u = \sum_i c_i f_i^{eq}$) the continuity equation as we expect from mass conservation

\begin{equation}
    \pdv{\rho}{t} + \pdv{\rho u}{x} = 0.
\end{equation}

For momentum conservation we have

\begin{equation}
    \sum_i c_i (f_i(x+c_i\Delta t,t+\Delta t) - f_i(x,t)) = 0.
\end{equation}
Applying it to Eq.\ref{eq:epsilont order}, thus multiplying for $c_i$ and summing over velocities we obtain
\begin{equation}
    \pdv{\rho u}{t} + \pdv{\sum_i c_i^2 f_i^{eq}}{x} = 0
\end{equation}
The second term is the pressure term $P^{(0)}$ as addressed in \cite{wolf2004lattice}. With our algorithm we have the equilibrium populations as in Eq.\ref{eq:feq_0}, obtaining 

\begin{eqnarray} \label{eq:pressure}
    P^{(0)} && = \sum_i c_i^2 f_i^{eq} \nonumber\\
    &&= \frac{\rho}{3} + \frac{1}{3}\frac{1}{\lambda_c + \lambda_s} ( 3 \lambda_c |\rho u| + 2\lambda_s \rho - \lambda_c \rho).
\end{eqnarray}
Thus simplifying, we obtain 

\begin{equation}
    \pdv{\rho u}{t} + \frac{\lambda_s}{\lambda_c + \lambda_s} \pdv{\rho}{x} + \frac{\lambda_c}{\lambda_c + \lambda_s} \pdv{|\rho u|}{x} = 0.
\end{equation}
Finally, at the order $o(\epsilon \Delta t)$ we have these two equations

\begin{align}
    \pdv{\rho}{t} + \pdv{\rho u}{x} & = 0, \\
    \pdv{\rho u}{t} + \frac{\lambda_s}{\lambda_c + \lambda_s} \pdv{\rho}{x} + \frac{\lambda_c}{\lambda_c + \lambda_s} \pdv{|\rho u|}{x} & = 0.
\end{align}

We can also look at the order $o(\epsilon^2 \Delta t^2)$, where we would have from mass conservation
\begin{equation}
    \pdv[2]{\rho}{t} + 2 \pdv{\rho u}{t}{x} + \pdv[2]{P}{x} = 0,
\end{equation}
and from momentum conservation we have
\begin{equation}
    \pdv[2]{\rho u}{t} + 2 \pdv{P}{t}{x} + \pdv[2]{\sum_i c_i^3 f_i}{x} = 0.
\end{equation}
Using $c_i^3=c_i$ and explicating the pressure term as in Eq.\ref{eq:pressure}, we obtain for $o(\epsilon^2 \Delta t^2)$

\begin{eqnarray} \nonumber
    \pdv[2]{\rho}{t} + 2 \pdv[2]{\rho u}{t}{x} + \frac{\lambda_s}{\lambda_s+\lambda_c} \pdv[2]{\rho}{x} + \frac{\lambda_c}{\lambda_s+\lambda_c} \pdv[2]{|\rho u|}{x} &  = 0, \label{eq:chapman_2_1}\\
    \pdv[2]{\rho u }{t} + 2 \qty( \frac{\lambda_s}{\lambda_s+\lambda_c} \pdv[2]{\rho}{t}{x} + \frac{\lambda_c}{\lambda_s+\lambda_c} \pdv[2]{|\rho u|}{t}{x}) + \pdv[2]{\rho u}{x} &  = 0. \nonumber
\end{eqnarray}

\nocite{*}

\bibliography{apssamp}

\end{document}